\documentclass{aa}  
\usepackage{graphicx}
\usepackage{natbib}
\usepackage{txfonts}
\begin{document}
   \title{PPM-Extended (PPMX) - a catalogue of positions and proper motions}
\author{S.~R\"{o}ser \inst{1} \and
E.~Schilbach  \inst{1} \and
H.~Schwan  \inst{1} \and
 N.V.~Kharchenko \inst{1,2,3}\and
 A.E.~Piskunov \inst{1,3,4}\and
  R.-D.~Scholz \inst{3} }

   \offprints{S.~R\"{o}ser}

\institute{Astronomisches Rechen-Institut, M\"{o}nchhofstra\ss{}e 12-14,
D--69120 Heidelberg, Germany\\
email: roeser@ari.uni-heidelberg.de, elena@ari.uni-heidelberg.de, schwan@ari.uni-heidelberg.de,
nkhar@ari.uni-heidelberg.de, \hfill \linebreak
apiskunov@ari.uni-heidelberg.de, 
\and
Main Astronomical Observatory, 27 Academica Zabolotnogo Str., 03680
Kiev, Ukraine\\
email: nkhar@mao.kiev.ua
\and
Astrophysikalisches Institut Potsdam, An der Sternwarte 16, D--14482
Potsdam, Germany\\
email: nkharchenko@aip.de, apiskunov@aip.de, rdscholz@aip.de
\and
Institute of Astronomy of the Russian Acad. Sci., 48 Pyatnitskaya
Str., 109017 Moscow, Russia\\
email: piskunov@inasan.rssi.ru
}
   \date{Received March 13, 2008; accepted May 28, 2008}

 
  \abstract
  {} 
   {We build a catalogue PPM-Extended (PPMX) on the ICRS system which is complete down to a well-defined limiting
   magnitude and contains the best presently available proper motions to be  suited for kinematical studies in
   the Galaxy.}
   {We perform a rigorous weighted least-squares adjustment of individual observations, spread over more than
   a century,   
   to determine mean positions and proper motions. The stellar content of PPMX is
    taken from GSC 1.2 supplemented by catalogues like ARIHIP, PPM and Tycho-2 at the bright end.
     All observations have been weighted according to their
   individual accuracy.  The catalogue
   has been screened towards rejecting false entries in the various source
   catalogues.}
   {PPM-Extended (PPMX) is a catalogue of 18,088,920 stars containing astrometric and 
   photometric information. Its limiting magnitude is about 15.2 in the GSC photometric system.
   PPMX consists of three parts: a) a survey complete down to R$_U$ = 12.8 in the magnitude
   system of UCAC2; b) additional stars of high-precision proper motions,
     and c) all other stars from GSC 1.2 identified in 2MASS.
   The typical accuracy of the proper motions is 2mas/y for 66 percent of the
   survey stars (a) and the high-precision stars (b), and about 10 mas/y for all
   other stars. PPMX contains photometric information from ASCC-2.5 and 2MASS.}
   {}
   \keywords{astrometry -- proper motions -- star catalogues}    
  \maketitle
%

\section{Introduction}

According to IAU Resolution B2 of the XXIIIrd General Assembly (1997), the Hipparcos
Catalogue \citep{1997yCat.1239....0E} is the primary realisation of the International
Celestial Reference System (ICRS)
at optical wavelengths.
Since then, it has been the aim of the astrometric community
to extend the Hipparcos system to higher star densities and fainter limiting
magnitudes. The first, and most important catalogue is Tycho-2 \citep{2000A&A...355L..27H},
based on observations of the Tycho experiment onboard the ESA-Hipparcos satellite.
The old-epoch observations of Tycho-2 were taken from new reductions \citep{1998AJ....115.1212U}
 of the
observations of the Astrographic Catalogue.
Tycho-2 contains about 2.5 million stars and is 90 percent complete down to
$V$ = 11.5. The Tycho B and V magnitudes were transferred to the Johnson system
in ASCC-2.5 \citep{2001KFNT...17..409K}.

In 2004, the UCAC2 catalogue \citep{2004AJ....127.3043Z} has been published.
It is based on a new full-sky survey made with a newly developed astrographic camera in the
years from 1998 to 2004. The first epochs are taken from yet unpublished remeasures of the AGK2 plates
and scans from the NPM and SPM plates,
also known as the Yellow Sky (YS). To a minor extent, old epochs come also from AC and smaller
astrometric catalogues of the last century.
UCAC2 is not yet
complete, the declination zone from about +40 degrees to the north pole
is still missing. UCAC2 contains some 48 million stars down to $R_U$ = 16 mag.
Proper motion errors are about 1-3 mas/yr for stars to 12th magnitude, and about 4-7
mas/yr for fainter stars to 16th magnitude.

The largest catalogue in the optical regime is USNO B1.0 \citep{2003AJ....125..984M}
with more than one billion objects. However, it should be noted that USNO B1.0
is not in the system of ICRS, it contains relative, not absolute proper
motions  \citep[see][]{2003AJ....125..984M}. A comparison of USNO B1.0 and
UCAC2 performed in the present work yielded systematic differences (in areas of
square degrees) of up to 5 mas/y in proper motions and up to
0.5 arcseconds at present-day epoch.

For kinematical studies in the Milky Way, more precisely in the closer
neighbourhood of the Sun, a catalogue of proper motions in the ICRS system
and with a
well-defined completeness limit is indispensable.
At Astronomisches Rechen-Institut, Heidelberg, we have therefore started
an initiative
to extend the Tycho-2 system to fainter magnitudes using all the
relevant astrometric observations of the last century, and perform
a rigorous least-squares adjustment to derive proper motions, instead
of determining proper motions from the diffences of positions from two epochs.

The first result of this effort is PPM-Extended (PPMX). This paper
summarizes the astrometric and photometric sources, from which PPMX is
build together with its properties. 
An important remark is appropriate
at this place. The aim of having a proper motion catalogue
of homogeneous accuracy and complete down to a well-defined
limiting magnitude in the visual regime (considerably fainter than Tycho-2)
is not achievable at present.
The main reason for this is the inhomogeneity of old-epoch observations
mainly of AC. So, PPMX is a compromise between having a bright completeness
limit and proper motions of homogeneous accuracy and a fainter
completeness limit with a more inhomogeneous distribution of the accuracies
of proper motions.  

In the next section we summarise the properties of the source catalogues which
enter PPMX, then we briefly describe the construction of the catalogue, and finish
with an overview of the properties of PPMX.

\section{Observational catalogues}~\label{obscat}

a) The Astrographic Catalogue

The Astrographic Catalogue project (initiated in 1887) was one of the largest enterprises
ever undertaken in astronomy. It resulted in about 10 million measurements of
rectangular plate coordinates for some 4.6 million different stars. AC is a double coverage
of the whole sky.
The observations and plate measurements were carried out by
20 observatories all over the world. Each observatory was responsible for a declination zone
on the sky.
The printed coordinates
have been independently digitised at the CDS, Strasbourg (partially), at the US Naval Observatory and
the Sternberg Institute
in Moscow. For PPMX we made use of information contained in each of the three data sets.
A full reduction of the AC in the Hipparcos system is given in \citet{2001AAS...19912904U}, called AC2000.2,
which presents averages of the individual measurements per star. 
We have attributed weights for the individual positions in the AC zones according to the
accuracy of the position measurements in the sources: these
weights were taken from \citet{TRC}.

b) The Guide Star Catalog

The Guide Star Catalog \citep[GSC 1.0,][]{1990AJ.....99.2019L}
was constructed to support pointing and target acquisition for the Hubble Space Telescope (HST).
For the construction of PPMX we used the GSC 1.2 \citep{2001AJ....121.1752M}, which saw
a considerable improvement of the astrometric quality due to a reduction of systematic errors as
functions of the stellar locus on the plate and on the stellar magnitude. Typical mean
errors of a coordinate in GSC 1.2 are about 0.3 arcsec \citep{2001AJ....121.1752M}.
GSC 1.2 contains some 25 million observations for about 18 million individual stars. As GSC 1.2
is not on the ICRS system (represented by Tycho-2), it was reduced to it via the
procedure described in sec.~\ref{sydif}.

c) 2MASS

The Two Micron All Sky Survey \citep{2006AJ....131.1163S}, 2MASS, is a complete Sky-Survey in
the J, H and K$_s$ bands performed in the years from 1997 to 2001. 2MASS is also a source of accurate
astrometric positions. Its 
astrometric accuracy is 
100 mas (1 $\sigma$)  relative to the Hipparcos reference frame for $K_s < $ 14 \citep{2006AJ....131.1163S}.
For the construction of PPMX we used the Point Source Catalog of some 471 million entries.

d) Tycho-2

The Tycho-2 catalogue \citep{2000A&A...355L..27H} is a compiled astrometric catalogue containing
positions and proper motions
as well as two-colour photometric data for the 2.5 million brightest stars in the sky. The modern-epoch
($\approx$ 1991.25) 
positions and magnitudes were obtained from the Tycho star mapper observations on the Hipparcos satellite.
The mean standard errors in astrometry range from about 4 mas at $V < 7.0 $ mag to about 90 mas for
stars fainter than $V = $ 12.0 mag. From Tycho-2 we exclusively used the 1991.25-epoch 
positions measured with the Hipparcos star mapper.  

e) CMC14

CMC14 \citep{2006yCat.1304....0C} is the latest catalogue from the Carlsberg Meridian Telescope, observed 
in the years 1999 to the
end of 2005. The data are obtained with a 2k$\times$2k CCD camera operating in a drift scan mode. Its
magnitude limit is 17 (Sloan r') and the positional accuracy is in the range from 35 to 100 mas.
The observations cover the declination band from -30 to +50 degrees, and the 
catalogue contains some 95 million stars.

f) The Bordeaux Carte-du-Ciel (CdC2000) catalogue

The Carte-du-Ciel project accompanied the Astrographic Catalogue. Each observatory was supposed
to map its zone in a single overlap with 3 exposures per plate, going 2 to 3 magnitudes fainter than
AC. This project was never completed.
The CdC2000 \citep{2006A&A...449..435R}
catalogue provides astrometric positions of the stars present on 512 Carte du Ciel plates archived
at Bordeaux Observatory covering a declination range $+11\hbox{$^\circ$ }\le\delta\le+18 \hbox{$^\circ$ }$.
It contains the positions of 344,781 stars down to $m_{pg} =$ 17.
The mean positional accuracy is 0.10-0.12$^{\prime \prime }$with mean epoch 1914.7.
This catalogue is particularly valuable for PPMX, because the limiting magnitude of the AC catalogue
in this declination zone is too bright.

g) UCAC2

The second US Naval Observatory  CCD Astrograph Catalog, UCAC2, \citep{2004AJ....127.3043Z} contains
positions and proper motions for some 48 million objects (mostly stars) and 
covers the sky area from -90$^\circ$ to +40$^\circ$ in declination, going up to +52$^\circ$ in some areas.
The accuracy of the positions obtained with the CCD Astrograph in the years 1998 to 2004 are 15-70 mas,
depending on magnitude, with estimated systematic errors of 10 mas or below.
UCAC2 is a compiled catalog of positions and proper motions referred to a standard epoch (J2000.0),
the original observations are not yet published. In this work we used UCAC2 with its given
astrometric position at the catalogue epoch J2000.0. 

h) ASCC-2.5

ASCC-2.5 \citep{2001KFNT...17..409K} is a catalogue of 2.5 million stars with proper motions in the Hipparcos system.
It is a compiled catalogue from sources like Tycho-2 - which overwhelmingly contributes - PPM, Hipparcos
and CMC 11. No astrometric data from ASCC-2.5 entered into PPMX, only the 
B and V magnitudes in the Johnson photometric system were taken instead of the $B_T$ and $V_T$ from Tycho-2. 

i) PPM

The PPM catalogue \citep{1991QB6.R66........} is a catalogue of positions and proper motions of 378,910 stars.
It contains the majority of all ground-based astrometric measurements (pre-Hipparcos) from the last century with
stars from the AC as first epochs. A rigorous weighted LSQ adjustment was applied to determine
positions and motions. In sec.~\ref{recon} we outline, how to incorporate the old PPM into
the least-squares adjustment for PPMX, and in  sec.~\ref{sydif} we describe the reduction of the PPM, which is
on the FK5 system, to the ICRS system.

j) ARIHIP

The ARIHIP Catalogue \citep{2001VeARI..40....1W} is a  combination of the results of the HIPPARCOS
astrometry satellite with ground-based data, and contains 90842 stars in total. The typical
mean error of an ARIHIP proper motion in the single-star mode \citep[see][]{1999A&A...347.1046W} is 0.89 mas/year. 
In the construction of PPMX, no attempt has been made to improve the astrometric information
contained in ARIHIP. Therefore the astrometric data from ARIHIP are simply copied for the 
stars present in PPMX. Note, that the more accurate
new reduction of the measurements of the Hipparcos satellite \citep{2007hnrr.book.....V} is not used in 
ARIHIP; also, they arrived too late to be incorporated in PPMX.

k) STARNET

The STARNET catalogue \citep{Starnet} has been constructed from GSC 1.0 and a new reduction
of the Astrographic Catalogue based on the keypunched dataset from the Sternberg Institute
in Moscow. STARNET contains some 4.3 million stars with accuracy of proper motions of
5 mas/y. Unfortunately, the original AC plate coordinates have not been published.
As STARNET
is not on the ICRS system (represented by Tycho-2), it was reduced to it via the procedure
described in sec.~\ref{sydif}.

The simple LSQ adjustment described in sec.~\ref{lsqadj} got only positions as input.
So, special care has been taken, that positions implicitly contained in some of the compilation
catalogues from above, did not enter the final catalogue more than once. If, for instance, STARNET
was used as a source catalogue, AC and GSC 1.2 positions have been reconstructed as described
in sec.~\ref{recon}. In this case,
no positions from AC2000.2 or from the published GSC 1.2 were used.
At this place, it seems appropriate to express a plea for publishing
original astrometric measurements, and not only derived and/or averaged coordinates.
This is especially important in the case of AC, where the full data sets
of x,y-coordinates reside at USNO and at the Sternberg Institute in Moscow.
Because of their enormous importance for proper motions these data sets must be
stored in astronomical data centres such as the CDS. The same holds for UCAC2, which
is a compiled catalogue, but the original observations from the USNO CCD-astrograph,
as well as the positions of the Yellow Sky are not yet published anywhere.

\subsection{Re-construction of the old epochs from PPM and STARNET}~\label{recon}

Both catalogues (PPM and STARNET) publish the full covariance matrix of the astrometric parameters
(i.e. mean epochs and mean errors of positions and proper motions). 
In this case we applied a method described by \citet{kopffetal} to equivalently replace
the positions and motions by two normal observations. 
Let $\overline{T}$ be the mean epoch, $\sigma_{p}$ and $\sigma_{p.m.}$ the mean errors of
position and proper motion per coordinate, respectively. Then 
$ w = \sigma_{p}^{-2} $ and $ w_{p.m.} = \sigma_{p.m.}^{-2} $ are the corresponding weights.
Both epochs of normal observations $T_1$ and $T_2$ and the corresponding weights
$w_1$ and $w_2$ are given from the equations
\begin{equation}
w  = \sum_{i=1,n}^{} w_i ,~w \overline{T} = \sum_{i=1,n}^{} w_i T_i,~ w_{p.m.}  = \sum_{i=1,n}^{} w_i (T_i-\overline{T})^2
\end{equation}
with $n=2$.
Suppose that one of the epochs, e.g. $T_2$ is given (known), then all the other
quantities are readily determined from the eqs.~(1). The positions at the epochs
$T_1$ and $T_2$ are calculated straightforwardly from the positions at
epoch 2000.0 and the proper motions.

In the case of a star in STARNET we know that $T_2$ is given by its GSC epoch, which is
published, and in the case of PPM the second epochs are either AGK3, CMC 1 to 4,
and FOCAT-S \citep{Bystrovetal} on the southern hemisphere.
In all cases the normal observations are easily derived, and, so, we can
profit from all the old catalogues which entered into the construction
of PPM.
The normal observations have been combined with the new positions from
Tycho-2, UCAC2, CMC14 and 2MASS in the LSQ adjustment
described in the next chapter.
\section{Construction of PPMX}
Four major steps were necessary to complete PPMX:
a) determining the star-list, b) cross-identifications of the
source catalogues with the star-list, c) reducing the individual
catalogues to the system of Tycho-2, and finally d) individual
LSQ adjustment.
\subsection{The star-list of PPMX}~\label{starlist}
The star-list of PPMX is a preliminary catalogue with
a mean position and a proper motion for each star. It is constructed in the following
way:
GSC 1.2 was taken as the starting point. It was cross-matched with itself to find stars
instead of observations. The cross-matching was on a 15$\times$15 arcsec window and gave the results
discussed in section ~\ref{acf}. All matches within 3 arcsec were attributed to an
individual star. These individual GSC 1.2 stars have been cross-matched with 2MASS in a 15$\times$15 arcsec
window. GSC 1.2 stars which could not be matched, did not make it into the star-list.
If a single star in  GSC 1.2 was
matched with a single star in 2MASS, this was rated a valid identification.
This allows to find stars with proper motions as large as 0.75 arcsec/y per
coordinate for a typical epoch difference of 20 years between GSC 1.2 and 2MASS
with all the risk of finding spurious proper motions. If more than one match occured within the
window, all GSC and 2MASS entries were gathered together and the GSC star pattern was shifted to
match the 2MASS pattern. For all the successful matches an average position at 2000.0 and a proper motion
was calculated, and this was called star-list 1.
Next, STARNET was cross-matched with star-list 1 in a 3$\times$3 arcsec window
at 2000.0.  All STARNET stars which could NOT be identified in star-list 1 were added to it to create
star-list 2. The same procedure was adopted for PPM and Tycho-2 to create star-list 3. Finally,  
AC2000.2 was matched with star-list 3 in a 3$\times$3 arcsec window at the AC2000.2 epoch.
All AC2000.2 stars which could NOT be identified were then cross-matched with 2MASS also in a 3$\times$3 arcsec window,
and those stars with matches were added to build the final star-list. 
   \begin{figure}[h!]
   \centering
   \includegraphics[bb=28 175 546 639,angle=-90,width=8cm,clip]{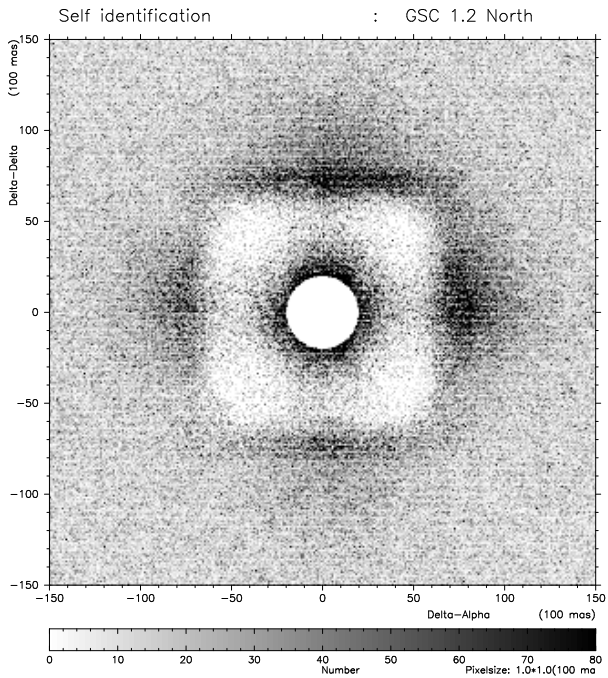}
      \caption{The two-point autocorrelation function of the northern part of GSC 1.2.
  The differences in right ascension and declination are given in units of 100 mas.
  The pixelsize is (100 mas)$^2$. For further explanation see text.}
         \label{autogscn}
   \end{figure}
%
 
\subsection{Cross-identifications}~\label{crossid}
All observational catalogues from section~\ref{obscat} with the exception of ARIHIP
were then cross-matched with the star-list in a 3$\times$3 arcsec window around the star-list position
at the epoch of the observation. If more than one match occured in this small window, the nearest neighbour
has been chosen. ARIHIP was only cross-matched after the LSQ adjustment for each star, and the
astrometric information was over-written.   

\subsection{Systematic differences to Tycho-2}~\label{sydif}

All observational catalogues have been cross-matched with Tycho-2 and 
individual differences in right ascension and declination have been calculated.
The individual differences were averaged in bins of 0.5$\times$0.5 degrees in
right ascension and declination, and the systematic differences in each bin
are determined via a 3$\times$3 bin moving average filter. This gives an effective area
of 2.25 square degrees, which, on average, contains 120 Tycho-2 stars. Only
in the case of PPM a bin size of one degree was taken because of the low spatial
density of PPM. After subtracting the so-determined spatial systematic differences, magnitude-dependent
differences within the magnitude range of Tycho-2 have also
been determined and were corrected for.

\subsection{The weigthed LSQ adjustment}~\label{lsqadj}
Weights $w_i$ are attributed to all n observations of a star according
to section~\ref{obscat}. Then the resulting covariance matrix of the unknowns
(mean position, proper motion) per coordinate is simply given by equation (1).
For the mean positions $\overline{x}$
and proper motions $\mu$ the following formulae hold
\begin{equation}
\overline{x} = \frac{\sum_{i=1,n}^{} w_i x_i}{\sum_{i=1,n}^{} w_i} ,~\mu  = \frac{\sum_{i=1,n}^{} w_i x_i (T_i-\overline{T})}{\sum_{i=1,n}^{} w_i  (T_i-\overline{T})^2}
\end{equation}

Automatic tests for unduly large scatter among the measurements (based on the $\chi^2$ sum)
and automatic elimination of obvious outliers 
(based on appropriately normalised individual residues) were implemented.
All stars having bad $\chi^2$ sums beyond a certain significance limit, but still not
showing obvious outliers, were marked as ''problem cases'' and got a 'P' flag
in the catalogue.

No attempt has been made to improve the astrometric quality of stars in ARIHIP, which
simply got their data copied from \citet{2001VeARI..40....1W}.

   \begin{figure}[h!]
   \centering
   \includegraphics[bb=28 175 546 639,angle=-90,width=8cm,clip]{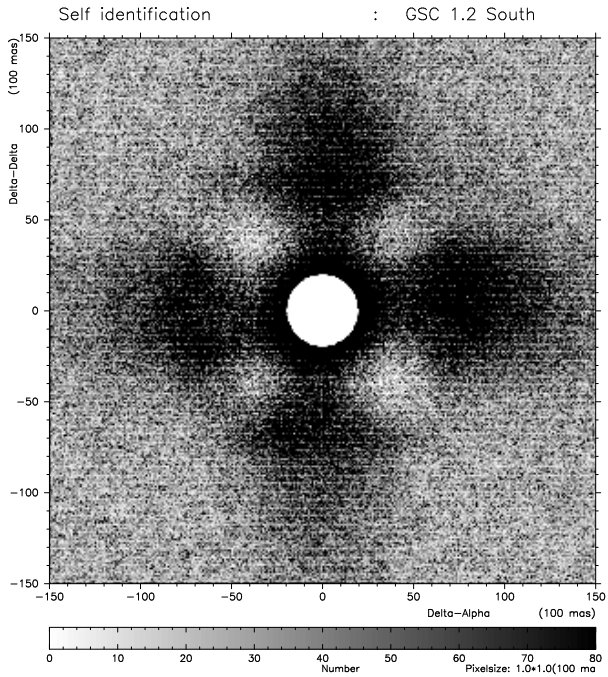}
      \caption{The same as Fig.~\ref{autogscn}, but for the southern part of GSC 1.2.}
         \label{autogscs}
   \end{figure}
%

\subsection{Autocorrelation functions}~\label{acf}

The plates that build the basis for GSC 1.2 partially overlap which may result in
up to 4 observations per star in this catalogue. We did not only rely on the
cross-matching in GSC 1.2, i.e using GSC 1.2 names, but performed a new one based
on coordinates only (described in section~\ref{starlist}).
A by-product of the large self-identification window is the detection of possibly spurious
entries in a catalogue. GSC 1.2 is based
on Schmidt plates and an automatic object detection has been carried out.
Spikes around brighter stars on Schmidt plates may give
rise to artefacts. The result of the self-identification of GSC 1.2 is the two-dimensional
 two-point autocorrelation function of GSC 1.2 on a 30$\times$30 arcsec scale
which is shown in Figs.~\ref{autogscn} and \ref{autogscs}, for
the northern, and the southern part of the sky, respectively. 
On a sphere randomly covered with single stars (one entry per star) the two-dimensional
autocorrelation function should be completely flat
outside the origin. A real excess on top of this flat distribution has to be 
attributed to physical double stars which would result in an increase radially symmetric around the
origin, given that the angular resolution of a catalogue is better than the separation of the
binary. Any deviation from radial symmetry has to be considered an artefact.

In Figs.~\ref{autogscn} and \ref{autogscs},
cross-matches inside a radius
of 2 arcsec are not plotted, thus generating an empty hole. If a match from
two different plates occured within 3 arcsec, this match was attributed to the same star.
Separations of less than 3 arcsec cannot be resolved in GSC.
One can see the wings of this distribution in both plots. We note that on the northern part based
on the Palomar Quick-V survey the authors of GSC 1.2 had already cleaned the catalogue
within a central square of 12$\times$12 arcsec. However, we still find hints to artefacts
from spikes on the Schmidt plates. This is more pronounced in the southern part, which is based on
deeper plates, and where
a different cleaning procedure could have been chosen.

   \begin{figure}[h!]
   \centering
   \includegraphics[bb=28 175 546 639,angle=-90,width=8cm,clip]{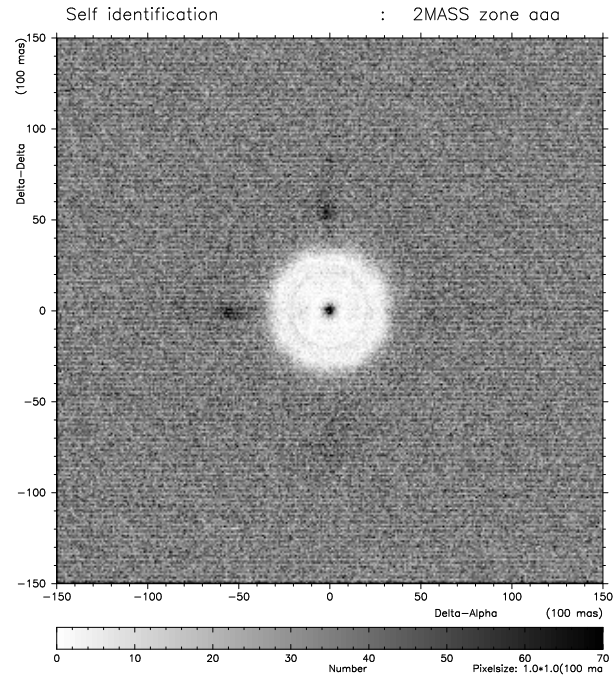}
      \caption{The autocorrelation function of 2MASS. Here, it is shown only for 
 the 5 million stars on the south polar cap
      (zone aaa). A full cross-matching of the .5 billion stars in 2MASS has not been
      performed.}
         \label{auto2ma}
   \end{figure}
%
For 2MASS a full autocorrelation of its about half a billion entries has not been
performed, only of the subset of 5~million stars  from the south polar cap (zone aaa). We assume
that this zone is representitative for 2MASS. 
The result of the autocorrelation of the subset is shown in Fig. \ref{auto2ma}. If this is
representative then 2MASS did an excellent job. The distribution is completely flat outside a radius
of 3 arcsec separation, except for a slight indication of spikes north and east of the origin.
In these comparisons we did not care about flags in 2MASS.
Note that the spikes are mostly inside a radius of 6 arcsec,
and the note for the 'prox' flag in 2MASS hints to be cautious if two entries have a distance less than 6 arcsec.
The spikes are more pronounced, if the central star is brighter than $K_s < 11$, and are practically absent
if  $K_s > 12$.

   \begin{figure}[h!]
   \centering
   \includegraphics[bb=75 39 533 771,angle=-90,width=9cm,clip]{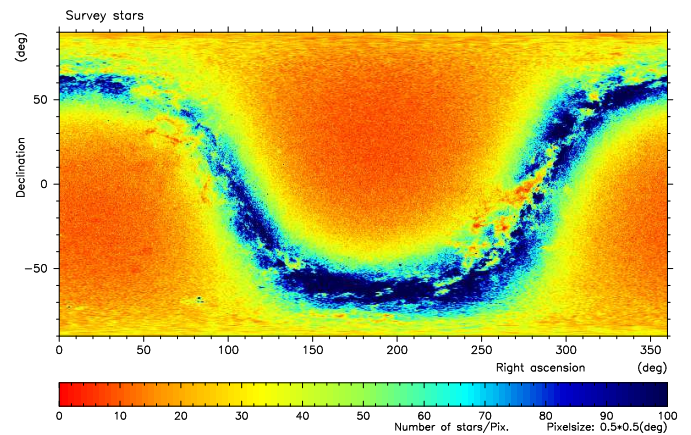}
      \caption{Number density of stars in the PPMX survey in equatorial coordinates.
       The bin size is (formally) 0.25 square degrees. No artefacts due to the lay-outs
       of the photographic surveys is seen here.}
         \label{n_ppmxsnow}
   \end{figure}
%

\section{Properties of PPMX}

The complete catalogue PPMX contains 18,088,920 stars. For convenience, we sub-divided the
catalogue into three disjunct subsets: the survey, additional high-precision stars and
all others.

\subsection{Selection of the different subsets}~\label{subsets}

PPMX uses sky surveys taken in different wavelength bands. The northern part of GSC 1.2, for instance, is observed
in the V-band, 2MASS in the infrared, and AC in the blue photographic range. Moreover, the different
AC zones have different limiting magnitudes mostly due to the efforts invested in manually measuring
the stars, which prevents us from simply taking the AC subset as a magnitude limited sample.
In addition, the photometric accuracy in AC is low and inhomogeneous. Also,
the photometric accuracy of GSC 1.2 is only about 0.3 mag \citep{2001AJ....121.1752M}, which would imply a rather soft boundary.
Neither can we adopt simply an infrared band from 2MASS, because this would lead to a low completeness
limit in the visual in order to take account of very cool or reddened stars, mainly at low galactic
latitudes. \citet{pisk2008} have introduced an artificial R$_J$ band constructed from the J and H
magnitudes in 2MASS. This R$_J$-band was calibrated towards the observed UCAC2 R$_U$  magnitudes.
We have taken the R$_J$-magnitude of 12.8 as the limiting magnitude of our subset called
"survey". This selection was guided by the requirement that the majority of stars in the survey
should be stars with observations in the AC to exploit their high accuracy of the proper motions.
Altogether there are 5,620,115 stars in the survey, and the are flagged 'S' in field 26 of the
catalogue. Among the survey stars, 3.7 million (66\%) have observations in AC, and therefore highly accurate
proper motions, whereas the remaining 1.9 million have typical mean errors of the proper motions of about 10 mas/y
(see Fig.~\ref{histoprop}).

Figure~\ref{n_ppmxsnow} shows the distribution of the survey stars on the sky plotted against
right ascension and declination.
This presentation
has been selected to show that there are no artificial enhancements or gaps related to the different zones
or plate borders
of the source catalogues.

   \begin{figure}[h!]
   \centering
   \includegraphics[bb=28 175 546 639,angle=-90,width=8cm,clip]{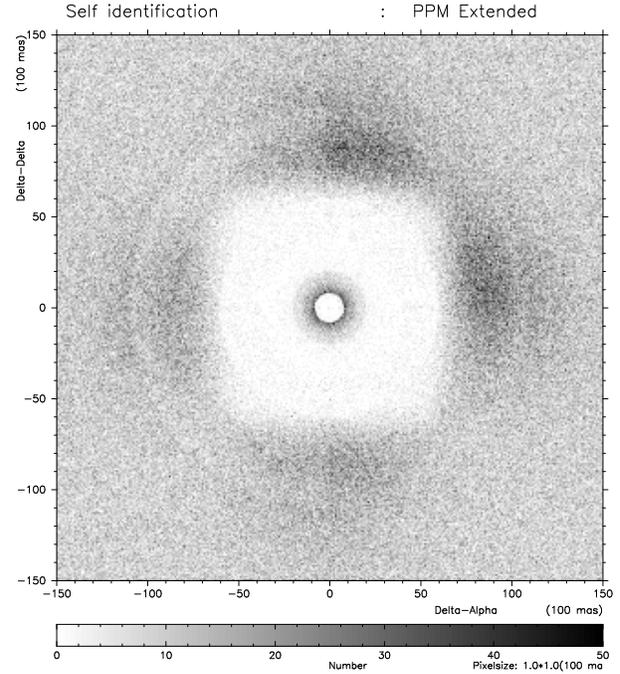}
      \caption{The autocorrelation function of PPMX.}
         \label{autoppmx}
   \end{figure}
%

Another 874,934 stars fainter than the survey limit have measurements in the AC and therefore
highly accurate proper motions ($\sigma_{p.m.}$ peaking at  about 3 mas/y per coordinate). They 
are gathered as "additional high-precision stars" flagged 'H' (see Fig.~\ref{histoprop}).
All other stars (11,593,871) get the flag 'O', their earliest observation epoch is from GSC 1.2.

For 57,334 stars in PPMX the determination of the R$_J$-magnitude was not possible, because
the magnitudes from 2MASS are contradictory. They are not contained in the survey even if they may
be bright enough to be members. Instead, they are flagged 'H' or 'O' in column 26 according
to the mean error of proper motions.

Users of PPMX may feel free to set other selection criteria than the one we used to define
our 'survey'.
   
\subsection{The autocorrelation function of PPMX}

Cross-identification of GSC 1.2 with 2MASS solved the majority of the problems with false
entries in both catalogues shown in Figs.~\ref{autogscn},~\ref{autogscs} and~\ref{auto2ma};
an additional cross-identification with CMC14 in the region between
-30 and +50 degrees declination made further improvement. We also deleted fainter companions closer
than 6 arcsec to a brighter star, if the fainter companion was only detected in GSC 1.2 as the earliest
epoch. Figure ~\ref{autoppmx} shows the 2-dimensional autocorrelation function of PPMX.
As can be seen, most of the spurious
entries from GSC 1.2 and 2MASS are gone. There is a remaining faint asymmetry north-east vs. south-west
of spurious stars whose origin could not be resolved. PPM and Tycho-2 have a better
spatial resolution than GSC 1.2 or 2MASS, therefore we have kept double stars
with flags 'P' and 'T' in column 27 even if their
separations are  smaller than 6 arcsec. Especially, the increase in density towards 1 arcsec separation
originates from physical double stars contained in Tycho-2.

   \begin{figure}[h!]
   \centering
   \includegraphics[bb=75 40 535 775,angle=-90,width=9cm,clip]{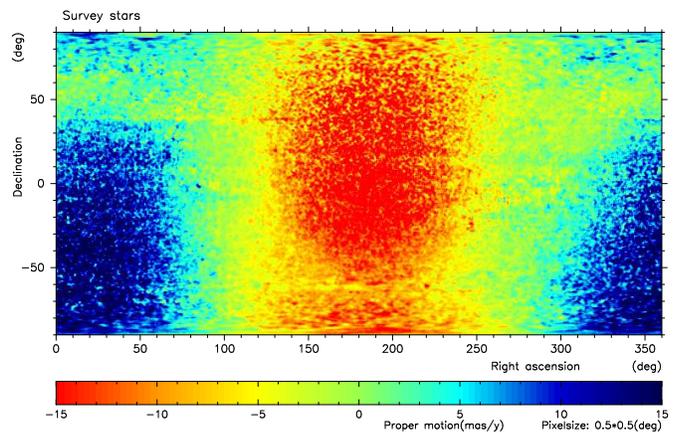}
      \caption{The proper motions in right ascension averaged in 0.25 square degree bins plotted over
      right ascension and declination for stars in the survey. The major effect seen here is the reflex of solar motion.
      The amplitude of the effects tells us that the bulk of the survey stars has distances between about 400 and
      500 pc from the sun.
      }
         \label{muealppmx}
   \end{figure}
%

\subsection{Astrometry in PPMX}~\label{astrom}

The individual catalogues have been reduced to the Tycho-2 system. So, remaining systematic deviations of,
e.g., the proper motion system of PPMX from that of Tycho-2 should not be expected. A comparison
showed that they are smaller than 1 mas/year on scales of a square degree or larger. There is, however,
an elegant way to test the proper motion system of a star catalogue. The proper motions of
a catalogue in an inertial reference system should only reflect the physical motions of the stars
in our Galaxy, i.e. the reflex of solar motion and the rotation of the Galaxy. Systematics
parallel to the axes of right ascension or declination must not appear, nor should one see features
representing the plate lay-out of a photographic survey.

\begin{figure}[h!]
   \centering
   \includegraphics[bb=75 40 535 775,angle=-90,width=9cm,clip]{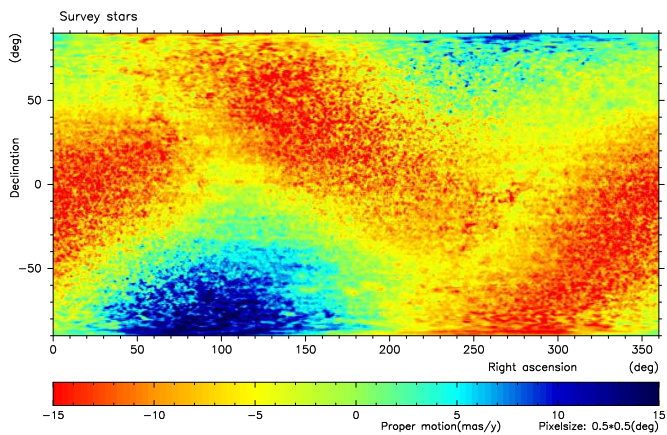}
      \caption{The same as Fig. ~\ref{muealppmx}, but for the proper motions in declination.}
         \label{muedeppmx}
   \end{figure}

Figs. \ref{muealppmx} and \ref{muedeppmx} show the  proper motions of the survey part of PPMX in the right ascension, declination plane. The
major feature to be seen in these plots is the solar reflex motion. The size of the effect is such that
it resembles a stellar sample with a typical distance of 400 to 500 parsec from the sun. 
There is an artificial depression of 1 to 2 mas/y in the proper motions in R.A. at declination +40
degrees, which is already inherent in Tycho-2.
The reason is unknown, it only correlates with the transition from the Potsdam/Hyderabad zone of AC to
the Helsingfors zone. 

For non-survey stars which, at the same time, are not in the high-precision subset, the plate pattern
of GSC 1.2 can bee seen in proper motion plots such as Fig.~\ref{muealppmx} at a 2 mas/y level,
especially in the northern declination range from the equator to +50 degrees. This plot is not shown
in the paper. The effect
resembles the
regions of plate overlap. As in these overlap regions the number density of stars is larger, too, it is unclear if 
the origin of these systematics
is a consequence of the larger number density or comes from still unresolved biases
in the GSC 1.2 plate reductions. The user is adviced to notice this effect in applications
based on stars with the 'O' flag in column 27.

%

The mean errors of the PPMX proper motions are determined in the LSQ adjustment
using eq. (1) with an a-priori error of unit weight
arbitrarily set to 1.
Due to the generally small
number of observations, hence low degree of freedom,
an a-posteriori determination of the error of unit weight from the residues 
per star leads to unreliable results. If, however, we used all residuals of all stars, we found
that the a-posteriori error of unit weight turned out to be about 5 percent smaller than the 
a-priori one, which proves that the error assignment to the observational catalogues was correct.

\begin{figure}[h!]
   \centering
   \includegraphics[angle=-90,width=8.8cm]{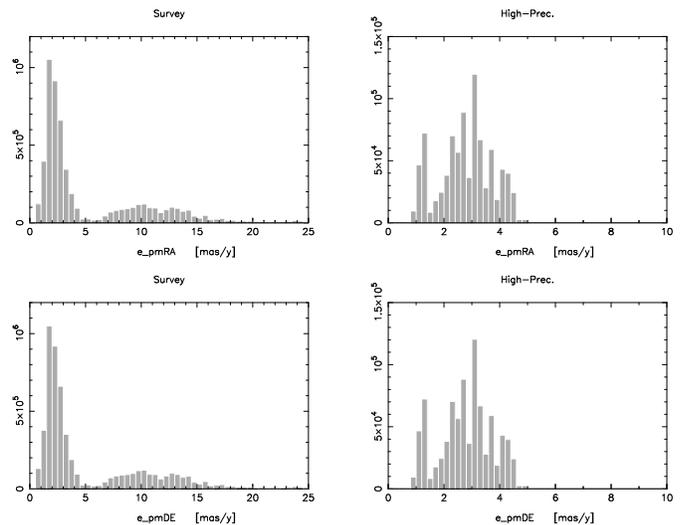}
      \caption{Histogram of the mean errors of proper motions in PPMX.
      The left panels show the distribution for the stars in the survey,
      the right panels for the high precision subset of PPMX.}
         \label{histoprop}
   \end{figure}

The distribution of the formal mean errors of the PPMX proper motions are shown in
Fig. \ref{histoprop}.

%

Here the distributions of the mean errors in proper motions of R.A. and Decl. are presented,
both for the 'survey' part (left panels) and the additional high-precision stars. 
Due to the selection criteria for the survey, the distribution is bi-modal . Two thirds of the stars
in the survey have AC observations and their mean errors peak between 1.5 and 2 mas/y.
The remaining stars got only observations
in 2MASS/UCAC2/CMC14 at the modern
epoch and GSC 1.2
at the old epoch and show a broad distribution between 5 and 15 mas /y. This mirrors
the distribution of the GSC 1.2 epochs ranging from the mid-seventies to the mid-eighties of the last
century. The distribution for the high-precision stars (right panels) shows
many individual peaks.  The reason for this is attributed to the Astrographic Catalog.
The individual zones of the AC have different accuracies of the measurements and, also, greatly vary in
their epochs. Later epochs give smaller epoch differences and hence a worse accuracy in the proper motions.
As frequently mentioned, this subset of high-precision proper motions is NOT
magnitude limited. All the remaining stars in the catalogue (flagged 'O') show the same distribution as the low-precision
part in the survey. Only a minority of the 'O' stars have significant (more than 3 times their mean errors) proper motions,
but all are published to create no selection effect.

\subsection{Photometry in PPMX}~\label{phot}

Photometric information in PPMX is copied from the source catalogues. All stars get a 'magnitude' in column 12
of the catalogue. This magnitude is taken from the source catalogue of column 27. For the overwhelming majority
these are GSC magnitudes. If the source is Tycho-2, the V$_T$ magnitude is given.
Also for  99.8\% of the stars 
photometry from 2MASS is available. For stars contained in the ASCC 2.5, B and V magnitudes in the
Johnson system are given (cols. 14 and 16). To construct a 
magnitude limited survey we have determined an artificial R$_J$ magnitude in the UCAC2 photometric
system from 2MASS photometry according to the method developed in \citet{pisk2008} (see section~\ref{subsets}).
This magnitude is given in column 13.

\begin{table}
\caption{Contents of the PPM-Extended catalogue. }
\label{tab_PPMX}
\setlength{\tabcolsep}{2pt}
\begin{tabular}{rlcl}
\hline
& Label&Units&Explanations\\
\hline
1  &name      &  ---    &  Name \\
2  &RAmas      &  mas   &  Right Ascension J2000.0, epoch 2000.0\\
3  &DEmas       & mas   &  Declination J2000.0, epoch 2000.0\\
4  &pmRA        & 0.01mas/yr & Proper Motion in RA$\cdot$cos(DE)\\
5  &pmDE        & 0.01mas/yr & Proper Motion in DE\\
6 & $T_{RA}$    &  0.01yr    &  Mean Epoch (RA) - 2000.00          \\
7 & $T_{DE}$    & 0.01yr    &  Mean Epoch (DE) - 2000.00           \\
8  &$e_{RA}$    & mas   & Mean error of Right Ascension\\
9  &$e_{DE}$    & mas   & Mean error of Declination\\
10 &$e_{pmRA}$  & 0.1mas/yr &  Mean error of pmRA$\cdot$cos(DE) \\
11 &$e_{pmDE}$  & 0.1mas/yr &  Mean error of pmDE\\
12 &$G$         & 0.001mag   &  $G$ magnitude from source catalogue  \\
 &  &			    &   (column 27)  \\
13 &$R_{J}$     & 0.001mag   & calculated $R_{J}$ magnitude (see text) \\
14 &$B$         & 0.001mag   & Johnson $B$ magnitude  from ASCC-2.5  \\
15 &$e_B$       & 0.001mag   &  Standard error of $B$ magnitude\\
16 &$V$         & 0.001mag   &  Johnson $V$ magnitude  from ASCC-2.5 \\
17 &$e_V$       & 0.001mag   &  Standard error of $V$ magnitude\\
18 &$J$         & 0.001mag   &  $J$ magnitude from 2MASS \\
19 &$e_J$       & 0.001mag   &  Standard error of $J$ magnitude\\
20 &$H$         & 0.001mag   &  $H$ magnitude from 2MASS \\
21 &$e_H$       & 0.001mag   &  Standard error of $H$ magnitude\\
22 &$K$         & 0.001mag   &  $K$ magnitude from 2MASS \\
23 &$e_K$       & 0.001mag   &  Standard error of $K$ magnitude\\
24 &$n_{obs}$     & ---   &  number of observations in the LSQ solution \\
25 &P         & ---   &  P flag. Bad LSQ fit\\
26 &sub         & ---   &  Subset flag: \\
   &            &       &   "survey (S)", \\
   &            &       &    "additional high-precision stars (H)" or \\
   &           &       &   "additional stars (O)"   \\
27 &I$_{sc}$    & ---   &  Index of source catalogue\\  
 &  &			    &  H stars from ARIHIP (Wielen et al.)   \\
&  &			    &  T stars from Tycho-2 (Hoeg et al.)   \\
&  &			    &  P stars from PPM (R\"oser and Bastian 1991)   \\
&  &			    &  S stars from STARNET (R\"oser 1996)   \\ 
&  &			    &  G stars (matching GSC 1.2 with 2MASS)   \\ 
&  &			    &  A stars (matching AC2000.2 with 2MASS)	\\
28 &Id        & ---   &  Identifier in the source catalogue \\
\hline
\end{tabular}
\end{table}

\section{The catalogue}~\label{res_sec}

The catalogue is subdivided into 24 declination zones of 7.5 degrees each. Stars in the zones are ordered
by increasing right ascension. The contents of the catalogue are described in Table~\ref{tab_PPMX}. The first column
gives the name of the star (in IAU convention HHMMSS.S$\pm$DDMMSS). The prefix "PPMX J" should be added
to identify a star in the catalogue.
There are 346 cases in which different stars would be attributed identical
names. Therefore, suffixes p (preceding) or f (following) are appended to distinguish between them.
Then follows the astrometric information (cols. 2 to 11), right ascension and declination at equinox and epoch
2000.0 in the ICRS system and the respective proper motions, as well as the full covariance matrix of these
quantities. This includes the mean epoch, mean error of the coordinates at mean epoch and mean errors of
proper motions. Columns 12 to 22 contain the photometric information as described in section~\ref{phot}.
Columns 24 and 25 present the number of observations used in the weighted LSQ solution, and a P flag
if the $\chi^2$ of the fit was unacceptable. There are 229,618 (or 1.6\%) cases with P flags in the catalogue.
Column 24 is left 'blank', if the stars originate from ARIHIP ('H' in column 27) or from Tycho-2 ('T' in column 27).
Column 25 describes, if a star belongs to the survey 'S', to high precision proper motion stars
'H' fainter than the survey limit, or the remaining stars 'O' (see~\ref{subsets}). Column 27 gives the source catalogues for the
star list(see section~\ref{starlist}), and column 28 the name of the star in the source catalogue.

This catalogue will be made available via the CDS, Strasbourg, France.

\subsection{Caveats}

Firstly, stars with high proper motions (HPM), having a total proper motion larger than about 200 mas/y and only 2 or 3 observations
(column 24 in the catalogue), may be spurious. Their proper motions are effectively based on two
epochs only, GSC1.2 as the first and 2MASS/UCAC2/CMC14 as the second. If one is interested in a particular
star, it is adviced to cross-match with USNO-A2.0, especially on the northern hemisphere.
Cross-matching of all the HPM stars was beyond the scope of the present paper, it will be done in the
near future. There are some 39000 stars with proper motions larger than 200 mas/y, but 16000 out of
them carry a 'P' (problem) flag in column 25.

Secondly, if a star in the catalogue has its nearest neighbour closer than 10 arcsec,
the fainter one of this pair may be spurious as seen in Fig.~\ref{autoppmx}. The pattern in Fig.~\ref{autoppmx} is a slight 
memory of the spurious entries in Figs.~\ref{autogscn} and~\ref{autogscs} from GSC 1.2, and~\ref{auto2ma} from 2MASS. The observed
overdensity in Fig.~\ref{autoppmx} gives rise to about 19000 of these cases.

Doubtful cases have been retained in PPMX. We publish them because, in a certain sense, they are 
measurements (or ``observations''). In order to verify in future what has been done, these
``observations'' have to be published.


\begin{acknowledgements}
Part of this work was supported by DFG grant 436 RUS 113
/757/0-2, and RFBR grants 06-02-16379 and 07-02-91566. This paper is based on
observations from the ESA Hipparcos satellite.
This publication makes use of data products from the Two Micron All Sky Survey,
which is a joint project of the University of Massachusetts and the
Infrared Processing and Analysis Center/California Institute of Technology,
funded by the National Aeronautics and Space Administration and the National Science Foundation.
This research has made use of the SIMBAD database,
operated at CDS, Strasbourg, France
\end{acknowledgements}

\end{document}